# NMR measurements and all-time Brownian movement with memory

Jana Tóthová and Vladimír Lisý

Department of Physics, Technical University of Košice, Slovakia

## ABSTRACT

Nuclear magnetic resonance (NMR), being nondestructive and noninvasive, is a widely used and very effective method to study random motion of particles in different systems, including the biological tissues. In the long time limit, when the particles are in the diffusion regime, the theoretical description of the NMR experiments is well developed giving proper interpretation of the measurements of normal and anomalous diffusion. The shorter-time dynamics is however correctly considered only within the standard memoryless Langevin description of the Brownian motion (BM). In the present work, by using the method of accumulation of phase shifts in the rotating frame, the attenuation function $S(t)$ of the NMR signal from an ensemble of spin-bearing particles in a magnetic-field gradient is expressed through the particle mean square displacement in a form applicable for any kind of stationary stochastic dynamics of spins and for any times. $S(t)$ is evaluated providing that the random motion of particles can be modeled by the generalized Langevin equation (GLE) with a colored random force driving the particles. The memory integral in this equation is the convolution of the particle velocity or its acceleration with a memory kernel related to the random force by the fluctuation-dissipation theorem. We consider three popular models of the BM with memory: the model of viscoelastic (Maxwell) fluids with the memory exponentially decaying in time, the fractional BM model, and the model of the hydrodynamic BM. In all the cases the solutions of the GLEs are obtained in an exceedingly simple way. The corresponding attenuation functions are then found for the free-induction NMR signal and the pulsed and steady-gradient spin-echo experiments. The known expressions valid for normal and anomalous diffusion are just limiting cases in the long time approximation. At shorter times the attenuation functions significantly differ from the classical formulas used to interpret these experiments. The results for the free-particle fractional BM compare favorably with experiments acquired in human neuronal tissues and with the observed subdiffusion dynamics in proteins. The effect of the harmonic trap is demonstrated by introducing a simple model for the generalized diffusion coefficient of the particle in a fractal environment.

# 1. INTRODUCTION

In the present contribution two fields are joined: the Brownian motion (BM) of particles in soft condensed matter and Nuclear Magnetic Resonance (NMR) as a tool of its observation. NMR has proven to be a very effective non-invasive method of studying molecular self-diffusion and diffusion in various materials and has a wide range of applications ranging from characterization of solutions to inferring microstructural features in biological tissues [1–13]. Brownian particles in some media often show interesting behaviors, such as a dynamics that significantly differs [14–17] from that predicted by the standard Einstein and Langevin theories [18–23]. It seems natural to apply NMR methods to study such unusual BM. However, as discussed in [24–29], the mathematical description of suitable NMR experiments in the literature is valid only for long times when the particles are in diffusion regime (normal or anomalous). An exception is the memoryless Langevin model [21], for which correct interpretations of the NMR experiments have been for the first time proposed in [30, 31]. Possible memory effects in the shorter-time particle dynamics are ignored or incorrectly taken into account.

The effect of diffusion on the NMR signal was incorporated by Torrey into the Bloch equations for the spin magnetization of an excited sample placed in a magnetic field [32]. By solving these equations the NMR signal is obtained as a product of the time evolution of the magnetization without the influence of diffusion and the diffusion suppression function $S(t)$. The function $S(t)$ can also be evaluated through the time-dependent resonance frequency offset in the frame rotating with the resonance frequency [1–4, 10, 30, 31, 33–41]. Evidently, the results based on the first approach do not go beyond the Einstein−Fick theory of diffusion. The second approach needs more attention. Most calculations of $S(t)$ within this approach are valid within the long-time approximation. This means that only experiments on spin-bearing particles undergoing diffusion in liquids or gases, when their mean square displacement (MSD) is recorded at times $t$ much larger than the characteristic frictional time of the particles [14], can be interpreted within the current theories. For shorter times, beyond the diffusion regime of the particle motion, these theories are inapplicable. In the works where $S(t)$ is successfully evaluated for all times, e.g. [30, 31, 42], the formulas for the attenuation of the NMR signal in a field gradient have been found in the frame of the standard Langevin theory of the BM [21]. This theory, however, has strong limitations. In fact, it is applicable only in the long-time limit when, however, it is not necessary since then the Einstein theory is valid. For shorter times it works only for Brownian particles with the density much exceeding that of the surrounding particles (such as dust particles in a gas) and is inappropriate for Brownian particles in a liquid [43–45]. With the aim to interpret NMR experiments at all times, new results based on a more general theory of the BM are thus highly desirable. In this paper, we first consider the previous attempts [30, 31, 38, 42] to describe the NMR experiments by using the generalized Langevin equation (GLE) with a memory integral [46, 47]. To do it, we begin with the discussion on the limitation of current calculations of $S(t)$ is given. Then we present a different way of finding $S(t)$ for an ensemble of stochastically moving spins in a magnetic-field gradient. The attenuation function will be calculated through the accumulation of the phase shifts in the rotating frame due to the particle displacements. Instead of the use of the particle positional autocorrelation function (PAF) as in [38], which is ill defined for unbounded random motion, $S(t)$ is expressed through the MSD. The obtained new formulas are independent on a model used for the description of the particle stochastic dynamics. It is only assumed that the studied random processes are stationary in the sense that the autocorrelation function of a fluctuating dynamical variable $x$ at times $t$ and $t'$ depends only on the time difference, $t - t'$: $\langle x(t)x(t')\rangle = \langle x(t-t')x(0)\rangle$ [10]. With this assumption the results are applicable to the BM with memory and at long times to normal and anomalous diffusion. An experiment will be described when the nuclear induction signal is read-out in the presence of a constant field gradient. Then the used method will be developed to the NMR pulse sequences based on the refocusing principle. Namely, the Hahn spin echo experiments with steady and pulsed field gradients will be described. Based on the works [24–26], a special section will be devoted to the derivation of the NMR attenuation signals assuming that the particle displacements can be described by the GLE, used in the literature to model the Brownian dynamics in viscoelastic (Maxwell) fluids [48, 49]. In the following sections we consider the influence of the hydrodynamic memory on the NMR signals [27], and finally the attenuation of the signals is studied within the popular fractional BM [29]. In all the cases the solution of the corresponding GLEs for free and harmonically trapped spin-bearing particles are obtained in an exceedingly simple way. We give several examples of the successful application of the presented theory to the descriptions of experiments, such as the measurements of water diffusion in neuronal tissue [42] and the internal dynamics in proteins [50].

## 2. LIMITATIONS ON THE CURRENT CALCULATIONS OF THE DIFFUSION SUPPRESSION FUNCTION

Following the paper [26], let us consider an experiment, in which the nuclear induction signal is observed in the presence of a steady magnetic-field gradient [38–41]. A liquid or gaseous system is placed in a strong magnetic field along the axis $x$ and after the 90° rf pulse the magnetization of an ensemble of spins is modulated by the field gradient. The influence of diffusion on the total magnetization determining the observed NMR signal is commonly expressed as [1–3, 10, 34–41]

$$S(t) = \langle \exp[i\phi(t)] \rangle = \left\langle \exp\left[ i\int_0^t \omega(\tau) d\tau \right] \right\rangle, \tag{1}$$

where $\omega(t)$ is the time-dependent resonance frequency offset in the rotating frame and the brackets $\langle ... \rangle$ mean the expectation value of the phase factor as a function of time [51]. In the literature [38], the relation $\omega(t) = \gamma_n g x(t)$ is used, where $g$ is the applied gradient strength, $\gamma_n$ is the nuclear gyromagnetic ratio, and $x(t)$ is the position of the spin after time $t$. It follows from Eq. (1) that

$$S(t) = \exp\left[-\frac{1}{2}\langle\phi^2(t)\rangle\right]. \tag{2}$$

The variable $\phi(t)$ varies randomly over the ensemble of spins. A sufficient assumption in obtaining Eq. (2) is the Gaussian distribution of $\phi(t)$, which is, according to the central limit theorem, a good assumption when dealing with very many randomly varying quantities [10]. For a non-Gaussian distribution, Eq. (2) holds approximately for a small $\phi$. A detailed derivation of this equation is given in [41]. Then the authors use the relation $\langle x(t)x(t+T)\rangle = D(2t+T-|T|)$, which, according to [40, 41], is obtained from the Einstein-Smoluchowski equation for 1D diffusion, $\langle x^2\rangle = 2Dt$. Here, $D = k_B T/\gamma$ is the diffusion coefficient determined by the temperature $T$ and the friction coefficient $\gamma$. From the above relation for the PAF one finds at $t = 0$ and $T > 0$ that $\langle x(0)x(T)\rangle = 0$. For 1D diffusion, an unambiguous quantity is the MSD, for which $X(t) = \langle[x(t)-x(0)]^2\rangle = 2Dt$ holds within the Einstein theory. Although in [40, 41] the correct result has been presented for $S(t)$, the used approach is applicable only for long times (much larger than the relaxation time $M/\gamma$ of the Brownian particle of mass $M$). It ignores the correlations between the particle positions that are present at shorter times and fails for stationary random processes considered in the present paper for all times. In Ref. [38], taking into account the stationarity, it has been obtained from Eq. (2)

$$S(t) = \exp\left[-\gamma_n^2 g^2 \int_0^t \langle x(t')x(0)\rangle(t-t')dt'\right]. \tag{3}$$

After substituting here $\langle x(t)x(0)\rangle \approx 2Dt$ as the PAF [38], the classical "textbook" expression for the diffusion suppression function was obtained,

$$S(t) = \exp\left[-\frac{1}{3}\gamma_n^2 g^2 Dt^3\right]. \tag{4}$$

Again, the result (4) is correct, but the way to obtain is not. It was used in [38] that $x(t) \approx x(0)$ when, however, the MSD $X(t)$ is zero instead of $2Dt$. Equation (4) was obtained from the incorrect formula (3) for $S(t)$ by using the incorrect expression for the PAF. The incorrectness of Eq. (3), aimed in [38] to describe the random motion of spins for all times $t$ was commented on in Ref. [24]. It can be seen also from the following consideration. Let the free spin-bearing particles exhibit normal diffusion at long times. If they are trapped in a harmonic well with elastic constant $k$, at the times much smaller than the characteristic time $\gamma/k$ the motion of the particles should not be affected by the trap so there is no reason that the influence of the trap would be reflected in $S(t)$. However, the MSD at $t \to 0$ behaves as $X(t) \approx k_B T t^2/M$ and the PAF for such bounded particles is known to be $\langle x(t)x(0) \rangle = k_B T/k - X(t)/2$ [43, 45]. Thus, in the short-time limit Eq. (3) becomes $S(t) \approx \exp[-\gamma_n^2 g^2 \langle x^2 \rangle t^2 /2]$ and depends on $k$ since $\langle x^2 \rangle \approx k_B T/k$. At long times, if $k \to 0$, the MSD becomes $X(t) \approx 2k_B T t/\gamma$. The substitution of $\langle x(t)x(0) \rangle$ in (3) then gives $S(t) \approx \exp[-\gamma_n^2 g^2 k_B T t^2 (1/2k - t/6\gamma)]$, whereas in the diffusion regime $S(t)$ should be determined by Eq. (4). For unbounded random motion the PAF cannot be used at all because in this case it is ill defined together with the quantity $\langle x^2 \rangle$ [52]. This holds for both the standard (memoryless) Langevin equation [21] describing the BM and for its generalizations that take into account the effects of memory [46, 47]. The latter case will be considered in below, where the GLE is applied to the description of the random motion of spins. Additional arguments will be given against the calculation of the attenuation function by using the PAF. Also, limiting expressions for the NMR signal will be given in the case when at long times spin ensembles with memory display anomalous diffusion.

## 3. THE ATTENUATION FUNCTION REVISITED

A quantity that should be used in the description of the influence of stochastic motion of spins on the NMR experiments is the well-defined and measurable MSD. The normal diffusion MSD at $t \to \infty$ tends to infinity as $2Dt$, which is not consistent with the approximation $\langle x(t)x(0) \rangle \approx \langle x(t)x(t) \rangle = 2Dt$ used in [38]. When the diffusion is anomalous, $X(t) = Ct^\alpha$, where $C$ is a temperature-dependent parameter, $\alpha = 1$ corresponds to normal diffusion with $C = 2D$, $\alpha < 1$ to sub-diffusion, and $\alpha > 1$ to super-diffusion [53, 54]. To capture all these cases and to obtain the attenuation function that would also be applicable to shorter times in the BM of spin-bearing particles, Eq. (3) should be modified as follows. The phase accumulation in the rotating frame in Eq. (1) must be calculated through the change of the phase during the time $t$, instead of the phase given by the spin position at time $t$, i.e., through the quantity $\Delta\omega(t) = \gamma_n g[x(t) - x(0)]$. This follows from the Larmor condition $\omega = -\gamma_n B$ relating the precessional frequency to the applied magnetic field $B = B_0 + g(t)x(t)$. For the random variable $x(t)$ we do not assume that $x(0) = 0$. Within the concept of accumulating phases [33], the deviation of the precessional phase of a spin at time $t$ from its phase at time $t = 0$ should be thus calculated as

$$\phi(t) = -\gamma_n \int_0^t \left[ B(x(t')) - B(x(0)) \right] dt' = -\gamma_n \int_0^t g\left[ x(t') - x(0) \right] dt'. \tag{5}$$

The difference from the approach used in [35] is that here the actual value of the magnetic field at time $t = 0$ is used, instead of its mean value $\langle B \rangle$, for which there is no reason. By using $\langle B \rangle$, $x(0)$ disappears from the above formula (if $x$ is a random process with $\langle x \rangle = 0$). As will be shown later, this modification of the approach [35] leads to very different results for the attenuation of the transverse magnetization $S = \langle \cos\phi \rangle$. While the formulas for the NMR signal obtained in [34, 35] and the subsequent papers (e.g., [3, 36]) are appropriate only in describing the diffusion of spins (i.e., their motion in the long-time approximation), our approach is equally suitable for the calculation of $S(t)$ at any times. The approximation in [34, 35] can be used to describe the diffusion regime of the particle motion but even for the stationary Markovian (memoryless) BM described by the standard Langevin theory it must be generalized. The standard Langevin equation can lead to notable corrections to the attenuation of the NMR signal due to diffusion if the frictional time of the Brownian particles is not much smaller than the characteristic time of the experiment (such as the time interval of the spin echo). The effect of memory in the particle dynamics can reveal itself if the memory function present in the GLE does not decay too fast (i.e., when the characteristic time of its decay is comparable to the time of experiment).

For the Gaussian random processes (or small $\phi$) we use Eq. (2), where now

$$\langle \phi^2(t) \rangle = \int_0^t \int_0^t dt'dt'' \langle \omega(t')\omega(t'') \rangle$$
$$= \frac{1}{2}\gamma_n^2 g^2 \int_0^t \int_0^t dt'dt'' \left[ X(t') + X(t'') - X(t''-t') \right]. \tag{6}$$

Since for stationary processes $X(t)$ is a symmetric function, one can use the following transformation:

$$\int_0^t \int_0^t dt'dt'' X(t''-t') = 2\int_0^t dt'(t-t')X(t'). \tag{7}$$

Equations (7), (6) and (2) then give the final simple result [24]

$$S(t) = \exp\left[ -\frac{1}{2}\gamma_n^2 g^2 \int_0^t t' X(t') dt' \right]. \tag{8}$$

This result is model-independent, applicable for any times and a character of the stochastic motion of spins, providing only that it is stationary and $\phi$ is Gaussian (or small). Most often, normal diffusion is observed in liquids and gases. Substituting $X(t) \approx 2Dt$ in (8), we return to the classical formula (4). The measured spectral line broadening due to diffusion (half width at half maximum) is $\omega_{1/2} \approx \sqrt{6}a^{1/3}$, where $a = \gamma_n^2 g^2 D/3$ [1]. At short times the motion of particles is ballistic [45], $X(t) \approx k_B T t^2 / M$, so that Eq. (8) gives

$$S(t) \approx \exp\left[-\frac{k_B T \gamma_n^2 g^2}{8M} t^4\right], \tag{9}$$

and $\omega_{1/2}^2 \approx 4\Gamma(5/4)\Gamma^{-1}(3/4)(k_B T \gamma_n^2 g^2 / 8M)^{1/2}$, where $\Gamma$ is the gamma function.

## 4. GENERALIZED LANGEVIN EQUATION WITH EXPONENTIALLY DECAYING NOISE AND THE NMR INDUCTION SIGNAL

As already discussed, the attenuation of the NMR signal within the model of standard Langevin equation for the BM has been correctly calculated in [30, 31, 42]. Here we will describe a more general case, which contains the Langevin theory as a special case. Recently [38], it has been proposed to describe the stochastic motion of spins in gases during the above considered NMR experiment using the GLE, in which the friction force is modeled by the convolution of the exponentially decaying memory kernel [48, 49, 54] $G(t) = (\gamma^2 / m) \exp(-\gamma t / m)$ with the particle velocity $\upsilon(t) = \dot{x}(t)$,

$$M\dot{\upsilon} + \int_0^t G(t-t')\upsilon(t')dt' = f(t). \tag{10}$$

Here, $f(t)$ is a stochastic force, $m \ll M$ is the mass of molecules in the surrounding medium and $\gamma$ is the friction coefficient proportional to the medium viscosity. By the fluctuation-dissipation theorem, the stochastic force describes colored noise, $\langle f(0)f(t)\rangle = k_B T G(t)$ [47]. Equation (10) is easily solved by the method presented in [54, 55] and used for a similar problem in [49]. This, in our opinion the most simple method of solving the GLE equations, goes back to the old works [56, 57] (rewritten and discussed in [58]), is as follows [54–61]. If we are interested in finding the MSD of the particle, $X(t) = \langle \Delta x^2(t) \rangle = \langle [x(t) - x(0)]^2 \rangle$, we have merely to replace $\upsilon(t)$ in (10) by $V(t) = dX(t)/dt$ and to substitute the stochastic force driving the particle with $2k_B T$. The equation of motion must be solved with the initial conditions $X(0) = V(0) = 0$. Obviously, also the condition $\dot{V}(0) = 2k_B T / M$ must hold. For more details see the above cited articles [54–61]. The new equation,

$$M\ddot{X}(t) + \int_0^t G(t-t')\dot{X}(t')dt' + kX(t) = 2k_B T, \tag{11}$$

is immediately solved by using the Laplace transformation. The solution for the velocity autocorrelation function $\langle \upsilon(t)\upsilon(0)\rangle$ determines the time-dependent diffusion coefficient $v(t) = \int_0^t \langle \upsilon(0)\upsilon(\tau)\rangle d\tau$,

$$v(t) = \frac{k_B T}{M} \left\{ \frac{\gamma}{m\zeta_-\zeta_+} - \frac{1}{\zeta_+ - \zeta_-} \left[ \left(1 - \frac{\gamma}{m\zeta_+}\right) e^{-\zeta_+ t} - \left(1 - \frac{\gamma}{m\zeta_-}\right) e^{-\zeta_- t} \right] \right\}, \tag{12}$$

where $\zeta_{-,+} = (\gamma/2m)(1 \mp \sqrt{1 - 4m/M})$. Note that in Ref. [38] the second term in {} was incorrectly obtained with the opposite sign. (It is easily seen, e.g., by taking the limit $m/M \to 0$, when the memory integral in (10) is replaced by $\gamma v(t)$ and one must get the result from the standard Langevin theory of the BM [21]. In this case $v(t)$ is

$$v(t) = \frac{k_B T}{\gamma} \left[ 1 - \exp\left(-\frac{\gamma}{M} t\right) \right], \tag{13}$$

while, by using $\zeta_+ \to \gamma/m$, $\zeta_- \to \gamma/M$, $\zeta_+ - \zeta_- \to \gamma/m$ (if $m/M \to 0$), and $\zeta_+\zeta_- = \gamma^2 (mM)^{-1}$ (for all $m$ and $M$), the result from [38] gives the sign + before the second term in the brackets. It is claimed in [38] that from $v(t)$ the PAF has been found having the form

$$\xi(t) = \frac{k_B T}{M(\zeta_+ - \zeta_-)} \left[ \frac{1}{\zeta_+} \left(1 - \frac{\gamma}{m\zeta_+}\right) e^{-\zeta_+ t} - \frac{1}{\zeta_-} \left(1 - \frac{\gamma}{m\zeta_-}\right) e^{-\zeta_- t} \right]. \tag{14}$$

Again, this is not correct for the following reasons. By integrating $2v(t)$ [49, 54], one obtains the MSD,

$$X(t) = \frac{2k_B T}{\gamma} t - \frac{2k_B T}{\gamma^2} (M - m) + 2\xi(t). \tag{15}$$

It has been used that $\zeta_+ + \zeta_- = \gamma/m$ and $\zeta_+\zeta_- = \gamma^2 (mM)^{-1}$. Equation (15) corrects also the solution of the GLE that has been obtained earlier in [30] and then used to calculate the attenuation of the NMR spin echo.

At $t \to \infty$ and $t \to 0$, respectively, one comes from (15) to the already used formulas $X(t) \approx 2k_B T t/\gamma$ and $X(t) \approx k_B T t^2/M$. The function $\xi(t)$ cannot be identified with $\langle x(t)x(0)\rangle$. Let us assume that $\xi(t) = \langle x(t)x(0)\rangle$. Simultaneously we have $X(t) = 2\langle x^2\rangle - 2\langle x(t)x(0)\rangle$, so that Eq. (15) can be rewritten to $X(t) = k_B T \gamma^{-1} t - k_B T \gamma^{-2}(M-m) + \langle x^2\rangle$. This relation does not correspond to the correct solution for $X(t)$, as it is seen already from the MSD long and short time limits or from the fact that $\langle x^2\rangle$ should be a constant. An evidently wrong expression $S(t) \propto \exp(\gamma_n^2 g^2 \kappa t)$ with $\kappa = k_B T M^2 \gamma^{-3}$ follows from (14) and (3) at $m \ll M$ and $\gamma t/m \gg 1$. If $\gamma t/m \ll 1$, $\xi(t)$ in the main approximation is $\xi(t) \approx k_B T M \gamma^{-2}$, so that at short times we have a decay of $S(t)$, as it should be, but at long times we have not, which again shows incorrectness of $S(t)$ found in [38]. By substituting (15) in (8), the new formula for $S(t)$ at long times reads

$$S(t) \approx \exp\left\{-\frac{k_B T \gamma_n^2 g^2}{3\gamma}\left[t^3 - \frac{3}{2\gamma}(M-m)t^2 + \frac{3M^3}{\gamma^3}\left(1 - \frac{3m}{M} + \frac{m^2}{M^2}\right)\right]\right\} . \tag{16}$$

This equation determines corrections to the classical result (4), providing the used GLE model is applicable. The linewidth broadening corresponding to (16) is given mainly by the law $\omega_{1/2} \sim (\gamma_n^2 g^2 k_B T / 3\gamma)^{1/3}$. If necessary, corrections to $\omega_{1/2}$ can be determined from (16) for a concrete system. Having a model for the friction coefficient $\gamma$ and the viscosity $\eta$, the temperature dependence of $\omega_{1/2}$ can be predicted. For example, in gases at high temperatures $\eta \sim \sqrt{T}$ [62]. Assuming the validity of the Stokes formula for $\gamma$, $\gamma = 6\pi\eta R$ ($R$ is the particle radius), this gives $\omega_{1/2} \sim T^{1/6}$ (with a correction that decreases with $T$ as $\sim T^{-1/6}$) instead of $\sim T^{-1/2}$ found in [38]. At low temperatures $\eta \sim T^{3/2}$ and $\omega_{1/2} \sim T^{-1/6}$. Note however that gases are systems where the memory in the particle dynamics is of low significance and the BM is very well described by the standard Langevin equation [44]. The considered GLE model seems to be applicable to viscoelastic fluids [48].

Equation (15) at $m \ll M$ becomes the MSD within the standard Langevin theory [21]. With this solution, Eq. (8) gives exactly the result obtained by Stepišnik [30], which at long times agrees with (16). In the case $4m > M$ the roots $\xi_{+,-}$ are complex and the solution (15) describes damped oscillations [35]. In the overdamped limit $4m \gg M$ one obtains from (15)

$$X(t) \approx 2D\left\{t + \frac{m}{\gamma}\left[1 - \exp\left(-\frac{\gamma t}{2m}\right)\cos\left(\frac{\gamma t}{\sqrt{mM}}\right)\right]\right\}. \tag{17}$$

This equation significantly differs from Eq. (30) in Ref. [30] obtained in the limit of large correlation times of the particles of mass $m$, surrounding the particles of mass $M$.

## 5. ATTENUATION OF THE SPIN ECHO SIGNAL

Modern NMR pulse sequences come from the simple refocusing principle of the spin echo developed by Hahn [63]. In this experiment, at time $t = \tau$ after the first 90° rf pulse at $t = 0$ the spin phases are inverted by a 180° pulse. Measurements of the echo signal amplitude at time $2\tau$ allow accurate determining of the diffusion coefficients of nuclear spins. During the experiment, a static magnetic field that creates macroscopic magnetization along the axis $x$ and a constant magnetic field gradient $g$ are applied. As in the case of the induction signal, we express the attenuation of the signal due to the stochastic motion of spins (2) through the accumulation of the changes of spin phases $\phi(t)$. Now we have

$$\langle \phi^2(t) \rangle = \gamma_n^2 g^2 \left\langle \left\{ \int_0^\tau [x(t') - x(0)] dt' - \int_\tau^t [x(t') - x(0)] dt' \right\}^2 \right\rangle. \tag{18}$$

The sign before the second integral accounts for the fact that at time $\tau$ all phases are inverted. Equation (18) can be again expressed through the MSD. After the averaging and use of the stationary condition one finds

$$\langle \phi^2(t) \rangle = \gamma_n^2 g^2 \int_0^\tau dt' (t' - 2\tau) X(t') + 2 \int_0^\tau dt' (2t' - t) X(t') + 2 \int_0^t dt' \int_0^\tau dt'' X(t' - t''). \tag{19}$$

Other equivalent forms of Eq. (19) are possible as well. In the special case of the Einstein diffusion we get from Eqs. (19) and (2) at $t = 2\tau$ the famous Stejskal-Tanner formula [64]

$$S(2\tau) = \exp\left[ -\frac{2}{3} \gamma_n^2 g^2 D \tau^3 \right]. \tag{20}$$

At arbitrary $t > \tau$

$$S(t) = \exp\left[ -\frac{1}{3} \gamma_n^2 g^2 D \left( t^3 - 6t\tau^2 + 6\tau^3 \right) \right]. \tag{21}$$

It is interesting that the maximum of the function $S(t)$ is not at the echo time $2\tau$ but earlier, at $t = \sqrt{2}\tau$. This has been for the first time obtained and experimentally verified in [40, 41].

Often the pulsed gradient method is used. Let the first gradient pulse begins at time $t = \tau_g$ after the 90° rf pulse and the second one at time $t = \Delta$ [10]. The 180° rf pulse is applied between these gradient pulses, the duration of each is $\delta$. The result for $\tau_g + \delta < \tau$ (up to the second rf pulse) is the same as for a steady gradient (Eq. (8) with $t = \delta$). Due to stationarity, after the second rf and gradient pulses the result of calculations also does not depend on $\tau_g$ (as distinct from [65]) and can be evaluated from

$$S(\delta, \Delta) = \exp\left\{ -\frac{1}{2} \gamma_n^2 g^2 \left[ \int_0^\delta dt' \int_0^\delta dt'' X(t'' - t' + \Delta) - 2 \int_0^\delta dt' (\delta - t') X(t') \right] \right\}. \tag{22}$$

This formula simplifies to the well-known relation [3] when the MSD is $X(t) = 2Dt$,

$$S(\delta,\Delta) = \exp\left[-\gamma_n^2 g^2 D\delta^2(\Delta-\delta/3)\right]. \tag{23}$$

As it will be shown below, for short-time pulses with $\delta \ll \Delta$ this expression can be used also to describe the signal from anomalously diffusing particles [3, 35]. Generally, however, this is not true. If $\delta = \Delta = \tau$ is substituted in (22), we obtain the damping of the signal at the echo time $2\tau$, in the case of the steady gradient, $S(\tau,\tau) = S(2\tau)$ which agrees with equations (2) and (19). The new exact expression for $S(t)$ within the GLE model described in the preceding section is obtained by substituting the MSD (15) in Eq. (22):

$$\frac{-\ln S(\delta,\Delta)}{(\gamma_n g)^2 D} = \delta^2\left(\Delta - \frac{\delta}{3}\right) + \frac{\gamma}{M}\frac{1}{\xi_+ - \xi_-}\left[\varphi(\xi_+) - \varphi(\xi_-)\right], \tag{24}$$

where

$$\varphi(z) = \frac{1}{z}\left(1 - \frac{\gamma}{mz}\right)\left\{-\frac{2\delta}{z} + \frac{2}{z^2}\left[1 - e^{-z\Delta} - e^{-z\delta} + \frac{1}{2}\left(e^{-z(\Delta+\delta)} + e^{-z(\Delta-\delta)}\right)\right]\right\}. \tag{25}$$

At long times Eq. (24) differs from (21) obtained for normal diffusion. The difference depends on the relation between the frictional time $M/\gamma$ and the experimental times $\delta$ and $\Delta$. In the main approximation for $\delta \gg m/\gamma$ and by using $m/M \to 0$, the result (24) converts to

$$\frac{-\ln S(\delta,\Delta)}{(\gamma_n g)^2 D} \approx \delta^2\left(\Delta - \frac{\delta}{3}\right) - 2\delta\left(\frac{M}{\gamma}\right)^2 + 2\left(\frac{M}{\gamma}\right)^3. \tag{26}$$

Formally this important result is the same as in the model described by the traditional Langevin equation for particles of mass $M$ [30]. However, while in the GLE model the correction to the attenuation function for normal diffusion can be notable, the long-time approximation in the Langevin model without memory requires $\delta \gg M/\gamma$, when the second and third terms in the right-hand side of Eq. (26) are very small. In its general form Eq. (24) significantly differs from the result in [30] for $S(t)$ obtained from the solution of the GLE (10), which was corrected in the preceding section.

## 6. ANOMALOUS DIFFUSION

The first attempt to describe the NMR experiments on systems displaying anomalous diffusion has been published by Jug [65]. Soon it has been shown [34, 35] that the results of [65] contradict the principle of time invariance and the correct expressions for the NMR spin echo attenuation $S(t)$ due to diffusion have been obtained. Assuming the MSD of the form $X(t) = Ct^\alpha$, and by using Eq. (8), it is easy to find $\langle \phi^2(t) \rangle$ and then $S(t)$ from (2) that generalizes the induction signal (4):

$$S(t) = \exp\left(-\frac{1}{2}\frac{\gamma_n^2 g^2 C}{\alpha+2} t^{\alpha+2}\right). \tag{27}$$

$S(t)$ in the Hahn echo experiment is readily obtained from (19) if, for symmetrical $X(t)$, the last integral is rewritten as

$$\int_0^\tau dt' \int_{-t'}^{t-t'} dx |x|^\alpha = \int_0^\tau dt' \left[\int_{-t'}^0 dx(-x)^\alpha + \int_0^{t-t'} x^\alpha dx\right] = \int_0^\tau dt' \left[\int_0^{t'} x^\alpha dx + \int_0^{t-t'} x^\alpha dx\right].$$

The final result of integration in (19) is

$$\frac{\langle \phi^2(t)\rangle}{C\gamma_n^2 g^2} = \frac{1}{(\alpha+1)(\alpha+2)}\left[t^{\alpha+2} - 2\tau^{\alpha+2} + (\alpha+2)(t-2\tau)(t^{\alpha+1} - 2\tau^{\alpha+1}) - 2(t-\tau)^{\alpha+2}\right]. \tag{28}$$

The minimum of $\langle \phi^2(t)\rangle$ is not at $t = 2\tau$ but earlier. It can be found from the equation

$$(\alpha+3)y^{\alpha+1} - 2(y-1)^{\alpha+1} - 2(\alpha+1)y^\alpha - 2 = 0, \quad y \equiv t/\tau > 1.$$

Only for $\alpha = 0$ one finds $y = 2$ (the echo time). For $\alpha = 1$ (normal diffusion) and 2 (ballistic motion), $y = \sqrt{2}$. For $\alpha \in (0, 2)$, $y$ changes from 2 to about $\sqrt{2}$ (with increasing $\alpha$ the minimum time decreases to $\sqrt{2}$ at $\alpha = 1$, then slightly increases and again decreases to $\sqrt{2}$ at $\alpha = 2$). From Eqs. (2) and (28), a simple formula follows at the spin echo time $2\tau$,

$$S(2\tau) = \exp\left[-2\gamma_n^2 g^2 C \frac{2^\alpha - 1}{(\alpha+1)(\alpha+2)} \tau^{\alpha+2}\right] \tag{29}$$

At $\alpha = 1$ we return to Eq. (20). In the case of pulsed gradient echo, it follows from (22) [35]

$$S(\delta,\Delta) = \exp\left\{-\frac{C\gamma_n^2 g^2}{2(\alpha+1)(\alpha+2)}\left[(\Delta+\delta)^{\alpha+2} + (\Delta-\delta)^{\alpha+2} - 2\Delta^{\alpha+2} - 2\delta^{\alpha+2}\right]\right\}. \tag{30}$$

This equation significantly differs from the result

$$S(\delta,\Delta) = \exp\left\{-\frac{1}{2}C\gamma_n^2 g^2 \delta^2\left[(\Delta-\delta)^\alpha + \frac{2\delta^\alpha}{2+\alpha}\right]\right\} \tag{31}$$

obtained in Ref. [42] and shows that the calculation of $\langle \Delta\phi^2(t)\rangle$ for the spin echo in [42], Eq. (43), the details of which are given in Appendix B, is not correct. When $\delta = \Delta = \tau$, from (30) we return to the result for steady gradient. When $\delta \ll \Delta$, for $X(\Delta) = C\Delta^\alpha$, (30) can be given the same form as for normal diffusion, $S(\Delta, \delta) \approx \exp[-\gamma_n^2 g^2 \delta^2 X(\Delta)/2]$ [3, 35].

## 7. NMR AND THE HYDRODYNAMIC BROWNIAN MOTION

It is a well-known fact that the description of the motion of particles in the normal ($\alpha = 1$) or anomalous diffusion regime assumes long times of observation – the times much larger than the relaxation (frictional) time of the particles [10] or the characteristic time of the loss of memory in the particle dynamics. The results of most theoretical studies in the literature correspond to such long times and are inapplicable to shorter times. This is rather surprising in view of a number of experimental investigations that have shown that the chaotic motion of Brownian particles in liquids at shorter times is not consistent with the classical theory by Einstein [19]. Also, an inconsistence with the Langevin theory of the BM [21] has been clearly demonstrated in a number of experiments, in the last decade particularly by the method of optical trapping [43–45, 66–70], that the Langevin equation very well describes the BM in fluids if it is modified by replacing the Stokes friction force with the Boussinesq-Basset "history force" [71, 72]. In such a generalized (hydrodynamic) theory of the BM the MSD at $t \to 0$ corresponds to the ballistic motion, $X(t) \sim t^2$, and at long times, in addition to the Einstein term $2Dt$, where $D$ is the diffusion coefficient, it contains important contributions due to the hydrodynamic memory. So, for a free particle this memory is displayed in the so-called long-time tails that slowly decrease to zero with the increase of time. These peculiarities should be reflected also in the attenuation function of the NMR signal. However, to our knowledge, for the hydrodynamic BM the corresponding calculations are absent. As to the calculations of $S(t)$ within the standard Langevin model [30], this theory has strong limitations: in fact, it is applicable only to Brownian particles with the density much exceeding that of the surrounding particles (such as the Brownian particles in a gas) and is inappropriate for the BM in liquids [44, 45]. Following [27], here we apply the hydrodynamic model of the BM that has been convincingly proven to well describe the BM in simple incompressible fluids. The above derived formulas for the attenuation of the NMR signals, expressed in a simple form suitable for calculations without reference to a concrete model of the particle stochastic dynamics (assuming only the Gaussian distribution of the random variables or small changes of the precessional phase of an individual spin during the experiment) will be used to evaluate the damping of the NMR relaxation signals due to the BM in the case of the steady-gradient and echo experiments.

Within the hydrodynamic theory the BM of particles is described by the GLE, in which the Stokes force is replaced by the Boussinesq force [71], naturally appearing as a solution of the linearized Navier-Stokes equations for incompressible fluids [73],

$$M^*\dot{\upsilon}(t) + \gamma\upsilon(t) + \int_{-\infty}^{t} G(t-t')\dot{\upsilon}(t')\mathrm{d}t' = f(t). \tag{32}$$

Here, $M^* = M + M_s/2$, $M_s$ is the mass of the fluid displaced by the particle of mass $M$, $\upsilon(t) = \dot{x}(t)$ is the particle velocity along one of the coordinate axes, say $x$, $\gamma$ is the friction coefficient, the memory kernel is here $G(t) = \gamma(\tau_R/\pi t)^{1/2}$, and $f(t)$ is the (colored) thermal noise force [54, 55, 59, 61]. The characteristic time $\tau_R = R^2 \rho/\eta$ for a spherical particle of radius $R$ in a liquid of density $\rho$ and viscosity $\eta$ determines how the memory in the particle dynamics is lost in time so that at long times, $t \gg \tau_R$, the particle reaches the (standard) diffusion regime. At shorter times, when $t$ is comparable or smaller than $\tau_R$, the behavior of the particle significantly differs from the diffusive one. The solution of Eq. (32) for the MSD has been obtained long ago [57] (for a review of this and later works see [58]). It can be given the form

$$\frac{X(t)}{2D} = t - 2\left(\frac{\tau_R t}{\pi}\right)^{1/2} + \tau_R - \tau^* + \frac{1}{\tau^*}\frac{1}{\lambda_2 - \lambda_1}\left[\frac{\exp(\lambda_2^2 t)}{\lambda_2^3}\text{erfc}(-\lambda_2\sqrt{t}) - \frac{\exp(\lambda_1^2 t)}{\lambda_1^3}\text{erfc}(-\lambda_1\sqrt{t})\right],$$
(33)

where $\tau^* = M^*/\gamma$, $\lambda_{1,2} = -(\tau_R^{1/2}/2\tau^*)[1 \mp (1-4\tau^*/\tau_R)^{1/2}]$, and erfc (.) is a complementary error function [74]. The coefficient $D$ is related to microscopic quantities as $D = (2\rho_p/9\rho)k_B T \tau_R/M = (2\rho_p/\rho + 1)k_B T \tau_R/9M^*$, where $\rho_p$ is the mass density of the particle. The solution (33) has been experimentally verified in a number of works, e.g., [66–69]. It significantly differs from the well-known solution of the standard Langevin equation [21] that is not appropriate for liquids but well corresponds to the BM in gases [44],

$$X(t) = 2D\{t - \tau[1 - \exp(-t/\tau)]\}, \quad \tau = M/\gamma.$$
(34)

The attenuation function (4) of the NMR induction signal due to the BM can be obtained exactly by integrating Eq. (33). By using the series representation of the error function [74] one finds $\int_0^t \exp(z^2 t')\text{erfc}(-z\sqrt{t'})dt' = z^{-2}[\exp(z^2 t)\text{erfc}(-z\sqrt{t}) - 1] - 2t^{1/2}/\sqrt{\pi}z$. With the help of this result the integral $\int_0^t t'X(t')dt'$ can be calculated by parts. Its long-time approximation for $t \gg \tau^*, \tau_R$ gives

$$S(t) \approx \exp\left\{-\frac{\gamma_n^2 g^2 D t^3}{3}\left[1 - \frac{12}{5}\left(\frac{\tau_R}{\pi t}\right)^{1/2} + \frac{3}{2}\frac{\tau_R - \tau^*}{t} + \left(\frac{9}{\pi}\right)^{1/2}\frac{(\tau_R - \tau^*)(\tau_R - 3\tau^*)}{\tau_R^2}\left(\frac{\tau_R}{t}\right)^{5/2}\right]\right\}.$$
(35)

This expression is more easily obtained from the long-time expansion of Eq. (33), if only the terms growing with $t$ are taken into account. A comparison with the result for the standard Langevin model (34),

$$S(t) = \exp\left\{-\gamma_n^2 g^2 D t^3 \left[\frac{1}{3} - \frac{\tau}{2t} + \left(\frac{\tau}{t}\right)^3 - \left(\frac{\tau}{t}\right)^2 \left(1 + \frac{\tau}{t}\right) \exp\left(-\frac{t}{\tau}\right)\right]\right\}, \tag{36}$$

shows that within the hydrodynamic model the classical result based on the Einstein theory of diffusion is reached much more slowly.

At very short times the motion is ballistic, $X(t) \approx k_B T t^2 / m$, and Eq. (4) gives Eq. (9), $S(t) \approx \exp[-k_B T \gamma_n^2 g^2 t^4 / 8m]$, with $m = M^*$ or $m = M$, depending on which of the two models is considered.

In the case of Hahn's echo, Eqs. (33) and (22) give for the attenuation function at long times

$$-\frac{\ln S(\delta, \Delta)}{\gamma_n^2 g^2 D} \approx \delta^2 \left(\Delta - \frac{\delta}{3}\right) - \frac{8}{15}\left(\frac{\tau_R}{\pi}\right)^{1/2}\left[(\Delta+\delta)^{\frac{5}{2}} - 2\Delta^{\frac{5}{2}} - 2\delta^{\frac{5}{2}} + (\Delta-\delta)^{\frac{5}{2}}\right]. \tag{37}$$

For the steady gradient we have to take $\Delta = \delta$ and obtain a correction to the Stejskal-Tanner result [64] at the echo time $2\delta$

$$S(\delta, \Delta \approx \delta) \approx \exp\left\{-\frac{2}{3}\gamma_n^2 g^2 D \delta^3 \left[1 - \frac{16}{5}(2^{1/2} - 1)\left(\frac{\tau_R}{\pi\delta}\right)^{1/2}\right]\right\}. \tag{38}$$

When $\delta \to 0$, the echo attenuation from (37) can be approximated by the formula $S(\delta, \Delta) \approx \exp[-\gamma_n^2 g^2 \delta^2 X(\Delta)/2]$. Here (cf. Eq. (33)), Eq. (37) at $\delta \ll \Delta$ determines an important correction to the result known for normal diffusion,

$$S(\delta \ll \Delta, \Delta) \approx \exp\left\{-\gamma_n^2 g^2 D \delta^2 \Delta \left[1 - 2(\tau_R / \pi\Delta)^{1/2}\right]\right\}, \tag{39}$$

and shows a difference from the attenuation within the standard Langevin theory of the BM, which at long times ($\delta \gg M/\gamma$) is [26, 30, 42]

$$-\ln S(\delta, \Delta) \approx (\gamma_n g)^2 D\left[\delta^2 (\Delta - \delta/3) - 2\tau^2 \delta\right]. \tag{40}$$

The obtained analytical results are illustrated by Figs. 1 – 3 that present numerical calculations [75] of the MSDs in the considered models and the corresponding attenuation functions. Figure 1 demonstrates the difference between $X(t)$ given by Eqs. (33) and (34), and the Einstein result $X(t) = 2Dt$. Figure 2 illustrates how the classical result from the Einstein diffusion theory, $S_E(t) = \exp(-\gamma_n^2 g^2 D t^3 / 3)$, is approached with increasing $t$ within the standard Langevin theory and in the theory of the hydrodynamic BM. The results from these theories, Eqs. (35) and (36), are plotted as $\ln S(t) / \ln S_E(t) = -3\ln S(t) / [(\gamma_n g)^2 D t^3]$, so that the classical result corresponds to unity. The time is expressed in dimensionless quantity $t/\tau$. For simplicity, in all figures it is assumed that the density $\rho$ of the liquid is equal to that of the particle ($\rho_p$), as for buoyant Brownian particles. In this case $\tau = 2\tau_R/9$ (it was used that $\tau = M/\gamma$, $\gamma = 6\pi\eta R$, and $M = 4\pi R^3 \rho_p /3$), and $\tau^* = M^*/\gamma = \tau_R/3 = 3\tau/2$. It is seen that the classical result is in the hydrodynamic model reached more slowly (basically as $\sim (\tau/t)^{1/2}$) than in the Langevin model (as $\sim \tau/t$). Even a more significant difference between the two models is demonstrated in Fig. 3 for the echo signal, when the classical result is $S(\delta,\Delta) = \exp[-\gamma_n^2 g^2 D \delta^2 (\Delta - \delta/3)]$. We assume that $\Delta \approx \delta$ and compare the Langevin model, Eq. (40), with Eq. (37). It can be concluded that if the diffusion coefficient is determined from experimental data, the error in its obtaining is time-dependent and can be large. For example, if $D$ would be measured in the experiment with steady gradient, even for $t/\tau$ as large as $10^3$ the difference between its value extracted by using the classical formula and the one from a more realistic hydrodynamic theory is predicted to be still about 10%.

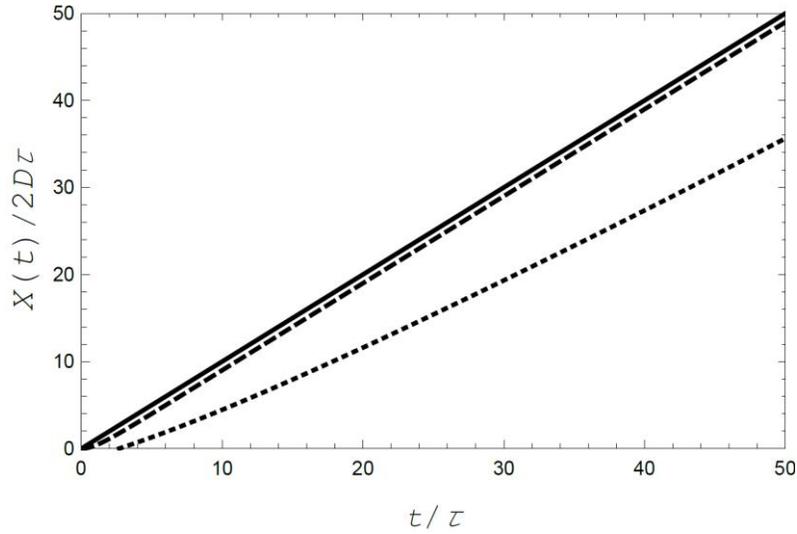

**Figure 1.** Demonstration of the differences between the exact result for the MSD within the hydrodynamic theory of the Brownian motion, Eq. (33) (dotted line), and the results from the standard Langevin (Eq. (34), dashed line) and Einstein (solid line) theories. It is assumed that the density $\rho$ of the liquid is equal to that of the particle ($\rho_p$). The MSDs are normalized to $2D\tau$, where $\tau = M/\gamma$ is taken as the unit of time.

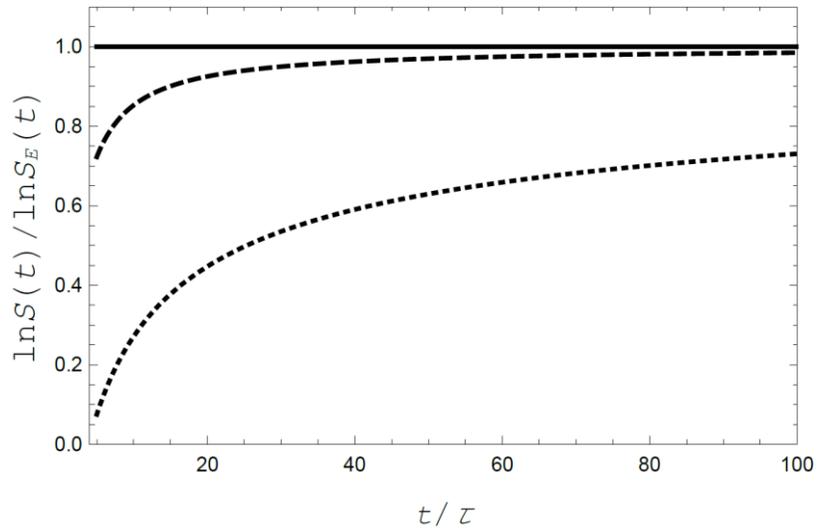

**Figure 2.** Comparison between the time dependence of the attenuation functions due to the Brownian motion in the steady-gradient experiment described within the Einstein theory (solid line), standard Langevin theory (Eq. (36), dashed line), and the hydrodynamic model (Eq. (35), dotted line). The numerical calculations correspond to the long-time behavior of $S(t)$ as described in the text, with normalization to the result based on the Einstein theory, which is thus presented as unity.

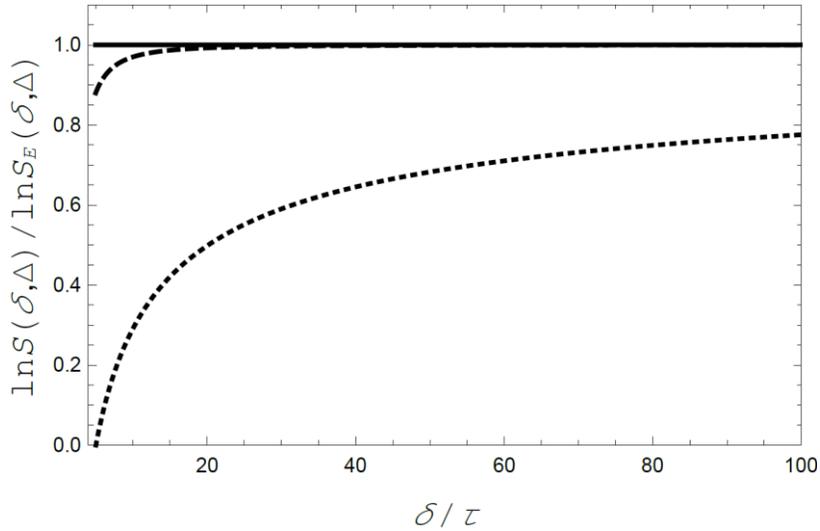

**Figure 3.** Comparison between the time dependence of the attenuation functions due to the Brownian motion in the steady-gradient spin-echo experiment described within the Einstein theory (solid line), the hydrodynamic model (Eq. (38), dotted line), and the standard Langevin theory (Eq. (40), dashed line). The numerical calculations correspond to long times as described in the text, with normalization to the result based on the Einstein theory, which is thus presented as unity.

## 8. NMR SIGNALS WITHIN THE MODEL OF FRACTIONAL BROWNIAN MOTION

In this section work we focus on a popular fractional BM model for the description of transport dynamics in complex fluids and different physical, chemical and biological systems [76–81], but also for open quantum systems, econophysics [82] and medicine [83]. In the condensed matter physics this model emerges naturally, e.g., in viscoelastic media, and interpolates between the standard Langevin equation (LE) with the white noise force and the purely viscous Stokes friction force, and a model with constant memory [81]. To describe the all-time dynamics of the particles and the possibility to probe it by the NMR methods, we obtain the solution of the GLE for this model in an exceedingly simple way. The Brownian particles can be free or trapped in a harmonic potential. The results for the MSD of the particles are then applied to calculate the attenuation function $S(t)$ for an ensemble of spins in a magnetic-field gradient. $S(t)$ is used in the form obtained above for any time and kind of the stochastic motion of spins. Although our work was mainly aimed to contribute to adequate interpretations of the NMR experiments on the BM of large particles in complex fluids, in the limit of free particles coupled to a fractal heat bath our results correct the description and give a favorable comparison with experiments acquired in human neuronal tissues [42]. Except for the description of diffusion of molecules in biological tissues (see also [4, 30]), the presented theory could be applied in interpretations of the NMR experiments on transport in heterogeneous systems such as the network of pores filled by a solvent [34, 35, 65], NMR measurements of the anomalous BM in microemulsions and polymers [3, 30, 84–89], chemical reaction kinetics [36, 90], distance fluctuations of particles in complex fluids, e.g., liquid crystals showing in the vicinity of the isotropic-to-nematic transition a power-law decay profiles in the relaxation behavior [91], dynamics of random-coil macromolecules in sufficiently concentrated solution (gels) [92, 93], anomalous BM phenomenon in crowded environments, such as living cells [37], and protein conformational dynamics [50, 94].

If a particle is moving in a trap modeled by a Hookean potential with elastic constant $k$, the GLE instead of Eq. (10) reads [47]

$$M\dot{\upsilon}(t) + \int_0^t G(t-t')\upsilon(t')dt' + k\int_0^t \upsilon(t')dt' = f(t), \qquad (41)$$

Now we will consider the memory kernel $G(t) = \gamma_\varepsilon \varepsilon t^{\varepsilon-1}$, $t > 0$, $1 > \varepsilon > 0$ [80]. Within the linear response theory the second fluctuation-dissipation theorem [47] relates $G(t)$ to the random force $f(t)$, $\langle f(t)f(0)\rangle = k_B T \varepsilon \gamma_\varepsilon t^{\varepsilon-1}$, $t > 0$. When $\varepsilon \to 0$, $G(t) \to 2\gamma\delta(t)$ and Eq. (41) becomes the standard memoryless Langevin equation with the white noise force and the purely viscous Stokes friction force $-\gamma\upsilon(t)$. The simple method of the solution of Eq. (10) is equally applicable here, giving the equation for the MSD $X(t)$,

$$M\ddot{X}(t) + \int_0^t G(t-t')\dot{X}(t')dt' + kX(t) = 2k_B T, \qquad (42)$$

which is immediately solved by using the Laplace transformation,

$$\tilde{X}(s) = \int_0^\infty X(t)e^{-st}dt = \frac{2k_B T}{Ms}\left[s^2 + \frac{\gamma_\varepsilon}{M}\Gamma(\varepsilon+1)s^{1-\varepsilon} + \frac{k}{M}\right]^{-1}, \tag{43}$$

where $\Gamma(z)$ is the gamma function [74]. Thus, in this way the exact solution of Eq. (41) [80], can be easily got in a few steps. The limiting cases of short and long times correspond to large and small $s$, respectively, so that if $\varepsilon < 1$ than at $s \to 0$ the main approximation of (3) will be $\tilde{X}(s) \simeq 2k_B T/ks$ and $X(t) \simeq 2k_B T/k$, if $t \to \infty$ [74]. At $s \to \infty$, $X(s) \simeq 2k_B T/Ms^3$ and $X(t) \simeq k_B T t^2/M$. Corrections to these expressions can be obtained from the exact representation of Eq. (3) in the time domain through the Mittag-Leffler functions $E_{\alpha,\beta}(y)$ [80],

$$X(t) = \frac{2k_B T}{M}\sum_{k=0}^\infty \frac{(-1)^k}{k!}(\omega t)^{2k} t^2 E^{(k)}_{1+\varepsilon,3+k-\varepsilon k}\left(-\varepsilon\gamma_\varepsilon \Gamma(\varepsilon)t^{1+\varepsilon}/M\right), \tag{44}$$

where $E_{\alpha,\beta}(y) = \sum_{j=0}^\infty y^j \Gamma^{-1}(\alpha j + \beta)$, $\alpha > 0$, $\beta > 0$, and $E^{(k)}_{\alpha,\beta}(y) = d^k E_{\alpha,\beta}(y)/dy^k$. It should be noted here that the short-time expansion of this exact solution for $X(t)$, as given in [80] (Eq. (41)), does not correspond to the solution of the standard Langevin equation (Eq. (41) in the limit $\varepsilon \to 0$) [95]

$$X(t) = \frac{2k_B T}{M}\frac{1}{\lambda_2 - \lambda_1}\left(\frac{1}{\lambda_1} - \frac{1}{\lambda_2} + \frac{\exp(\lambda_2 t)}{\lambda_2} - \frac{\exp(\lambda_1 t)}{\lambda_1}\right), \tag{45}$$

with $\lambda_{1,2}$ being the roots of the equation $\lambda^2 + \gamma\lambda/M + k/M = 0$, $2\lambda_{1,2} = -\omega_M[1 \mp (1-4\omega^2\omega_M^{-2})^{1/2}]$, where $\omega^2 = k/M$, $\omega_M = \gamma/M$ and $\lambda_1\lambda_2 = \omega^2$. In [80], Eq. (41), the term $\sim t^4$ in $\{.\}$ should be $(1/12)(\omega_M^2 - \omega^2)t^4$, i.e., the term containing $\omega_M$ is missing. Another way to show that the expansions in [80] should be completed is to consider the limit of the free Brownian particle (when $\omega = 0$). The velocity autocorrelation function in this case must be $\langle v(0)v(t)\rangle = (k_B T/M)\exp(-\omega_M t)$, which does not agree with Eq. (42) [80]. The expansion that gives the correct simplification to the solution of the standard LE is for $C_v(t) = \langle v(0)v(t)\rangle/\langle v^2\rangle$

$$C_v(t) = 1 - \frac{\gamma_\varepsilon}{M(1+\varepsilon)}t^{1+\varepsilon} + \left(\frac{\gamma_\varepsilon}{M}\right)^2 \frac{\Gamma^2(1+\varepsilon)}{\Gamma(3+2\varepsilon)}t^{2+2\varepsilon} - \frac{1}{2}(\omega t)^2 + \ldots, \tag{46}$$

and the MSD should be

$$\frac{M}{k_B T}X(t) = t^2 - 2\frac{\gamma_\varepsilon}{M}\frac{\Gamma(1+\varepsilon)}{\Gamma(4+\varepsilon)}t^{3+\varepsilon} + 2\left(\frac{\gamma_\varepsilon}{M}\right)^2 \frac{\Gamma^2(1+\varepsilon)}{\Gamma(5+2\varepsilon)}t^{4+2\varepsilon} - \frac{\omega^2 t^4}{12} + \ldots, \tag{47}$$

In the case of a free particle, the solution of the GLE takes the form [80]

$$X(t) = \frac{2k_B T}{M} t^2 E_{\varepsilon+1,3}\left(-\frac{\gamma_\varepsilon}{M}\Gamma(\varepsilon+1)t^{\varepsilon+1}\right). \tag{48}$$

This equation was used by Cooke et al. [42] in the calculation of the attenuation function $S(t)$ of the NMR signals in diffusion-weighted experiments on human neuronal tissues. Below it will be shown that this calculation must be corrected in several points. As a consequence, the final result for the spin-echo signal proposed in [42] must be changed. $S(t)$ will be also obtained for bounded spin-bearing particles, whose MSD is described by Eq. (43) and its short- and long-time approximations in the time domain. The basic formulas used to calculate $S(t)$ are presented in the preceding sections.

In Ref. [42], in the case of the steady gradient, the solution (48) was used to calculate $\langle\phi^2(t)\rangle$ for $S(t)$ in Eq. (2) from the expression

$$\langle\phi^2(t)\rangle = 2\gamma_n^2 g^2 \int_0^t\int_0^t t_1 t_2 \langle\dot{x}(t_1)\dot{x}(t_2)\rangle dt_1 dt_2, \tag{49}$$

which is correct (although more complicated than that used in our Eq. (8)) but its derivation is wrong. Cooke et al. [42] apply the equation $\phi^2(t) = 2\int_0^t \phi(t_1)\dot{\phi}(t_1)dt_1$. They get $\phi(t) = -\gamma_n \int_0^t\int_0^{t_1} dt_1 dt_2 \dot{x}(t_1)g(t_2)$, which is not correct. In its derivation by integration by parts from the equation $\phi(t) = \gamma_n \int_0^t x(t_1)g(t_1)dt_1$ they impose the "rephasing condition" $\int_0^t g(t')dt' = 0$, which evidently does not hold for the steady gradient. Nevertheless, when using the correct expression $\phi(t) = \gamma_n g t x(t) - \gamma_n g \int_0^t \dot{x}(t_1)t_1 dt_1$ (for a constant gradient), the correct result (49) is obtained. This is because the term $\gamma_n g t x(t)$ in $\phi(t)$ does not contribute to $\langle\phi^2(t)\rangle$. However, the calculation of $\langle\phi^2(t)\rangle$ for the spin echo in [42], Eq. (43), the details of which are given in Appendix B, is wrong. It is seen from the long-time dependence of $\langle\phi^2(t)\rangle$, which must be the same as that for anomalous diffusion with the MSD $\langle X(t)\rangle = Ct^\alpha$. The correct $\langle\phi^2(t)\rangle$ for this case [35] (see Eq. (30) with $\alpha = 1 - \varepsilon$) significantly differs from the result in [42]. In the calculation of $\langle\Delta\phi^2(t)\rangle$ for the spin echo they consider four intervals, viz., $(0, \delta)$, $(\delta, \Delta)$, $(\Delta, \Delta + \delta)$, $(\Delta + \delta, 2\tau)$, and evaluate the contributions to $\langle\phi^2(t)\rangle$ separately for all the intervals. In spite of the fact that in the second and fourth interval the gradient is absent, the contributions are found to be $\sim g$. The summation then gives the incorrect result, Eq. (43) [42]. Here, instead of (49), we use a more simple expression (8). With this equation and the solution (48), the attenuation function is straightforwardly calculated. By using the expansion of $E_{\alpha,\beta}(y)$ shown after Eq. (44), the $t \to 0$ behavior of $S(t)$ is given mainly by the formula $S(t) \approx \exp[-\gamma_n^2 g^2 k_B T t^4 / 8M]$, Eq. (9), which is the same as in the standard Langevin and other models. The correction to $\langle \phi^2(t) \rangle$ in the considered model is given by the equation

$$\langle \phi^2(t) \rangle = \frac{2\gamma_n^2 g^2 k_B T}{M} \left( \frac{t^4}{8} - \frac{\gamma_\varepsilon \Gamma(\varepsilon+1)}{M(\varepsilon+5)\Gamma(\varepsilon+4)} t^{\varepsilon+5} + ... \right). \tag{50}$$

At $\varepsilon = 0$ it agrees with the standard Langevin model, for which the MSD is $X(t) = 2D\{t - (M/\gamma)[1 - \exp(-\gamma t/M)]\}$.

To obtain the long-time limit at $\gamma_\varepsilon \Gamma(\varepsilon+1) t^{\varepsilon+1} / M \gg 1$, one can use [78–80] $E_{\alpha,\beta}(-z) \sim [z\Gamma(\beta-\alpha)]^{-1}$, $z > 0$, and the properties of the gamma function [74] $\Gamma(z+1) = z\Gamma(z)$ and $-z\Gamma(z)\Gamma(-z) = \pi \csc \pi z$. The MSD is then approximated by the formula $X(t) \approx C_\varepsilon t^{1-\varepsilon}$ with $C_\varepsilon / 2k_B T = [\gamma_\varepsilon \Gamma(2-\varepsilon)\Gamma(1+\varepsilon)]^{-1} = [\gamma_\varepsilon (1-\varepsilon)\pi\varepsilon]^{-1} \sin \pi\varepsilon$. Thus, for the steady gradient at long times

$$\langle \phi^2(t) \rangle \approx \frac{C_\varepsilon \gamma_n^2 g^2}{3-\varepsilon} t^{3-\varepsilon}. \tag{51}$$

In the case of the pulsed-gradient echo, the decrease of the NMR signal at any time after the second gradient pulse will be from Eq. (22)

$$\langle \phi^2(\delta, \Delta) \rangle \approx \frac{C_\varepsilon \gamma_n^2 g^2}{(2-\varepsilon)(3-\varepsilon)} \left[ (\Delta+\delta)^{3-\varepsilon} + (\Delta-\delta)^{3-\varepsilon} - 2\Delta^{3-\varepsilon} - 2\delta^{3-\varepsilon} \right]. \tag{52}$$

This equation corresponds to the result found in [35] but significantly differs from that by Cooke et al., Eq. (43) [42], where $\langle \phi^2(\delta, \Delta) \rangle \propto \delta^2 [(\Delta-\delta)^{1-\varepsilon} + 2\delta^{1-\varepsilon} / (3-\varepsilon)]$.

For the steady-gradient experiment we obtain from Eqs. (9) and (8) with the MSD (47)

$$\langle \phi^2(t) \rangle = \frac{\gamma_n^2 g^2 k_B T}{M} \left( \frac{t^4}{4} - \frac{2\gamma_\varepsilon \Gamma(1+\varepsilon)}{M(5+\varepsilon)\Gamma(\varepsilon+4)} t^{5+\varepsilon} + \left( \frac{\gamma_\varepsilon}{M} \right)^2 \frac{\Gamma^2(1+\varepsilon)}{(3+\varepsilon)\Gamma(5+2\varepsilon)} t^{6+2\varepsilon} - \frac{\omega^2}{72} t^6 + ... \right). \tag{53}$$

The expansion is done up to the term $\sim t^6$ corresponding to the fact that the particle begins to "feel" the trap. At $\varepsilon = 0$ this formula agrees with that from the solution (45) of the standard LE for a trapped particle, and at $\omega = 0$ we have the result for a free particle.

The echo attenuation obtained from Eq. (22) is at short times

$$\langle \phi^2(\delta, \Delta) \rangle = \frac{\gamma_n^2 g^2 k_B T}{M} \left( \frac{1}{12} \varphi_4 - 2 \frac{\gamma_\varepsilon}{M} \frac{\Gamma(1+\varepsilon)}{\Gamma(6+\varepsilon)} \varphi_{5+\varepsilon} + 2 \left( \frac{\gamma_\varepsilon}{M} \right)^2 \frac{\Gamma^2(1+\varepsilon)}{\Gamma(7+2\varepsilon)} \varphi_{6+2\varepsilon} - \frac{\omega^2}{360} \varphi_6 + ... \right), \tag{54}$$

with $\varphi_\alpha = (\Delta+\delta)^\alpha + (\Delta-\delta)^\alpha - 2\Delta^\alpha - 2\delta^\alpha$. At $\delta \ll \Delta$ it holds $\varphi_\alpha \approx \alpha(\alpha-1)\Delta^{\alpha-2}\delta^2$ and one obtains $\langle\phi^2(\delta,\Delta)\rangle = (\gamma_n g\delta)^2 X(\Delta)$, with $X(t)$ from (47). It is easy to return to the solution based on the standard Langevin theory for both the trapped ($\varepsilon = 0$) and free particles ($\varepsilon = 0$, $\omega = 0$).

The long-time approximation needs more attention. In the work [80], Eq. (48), the MSD obtained for $t^{1-\varepsilon} \gg \gamma_\varepsilon \Gamma(1+\varepsilon)/M\omega^2$,

$$X(t) \approx \frac{2k_B T}{M\omega^2}\left[1 - \frac{\varepsilon\gamma_\varepsilon}{M\omega^2}t^{-1+\varepsilon}\right], \tag{55}$$

cannot be used to get the $\omega \to 0$ limit and the standard Langevin limit with a trap ($\varepsilon = 0$) gives only the main approximation $X(t) \approx 2k_B T/M\omega^2$. To obtain the long-time behavior valid also for the vanishing trapping force one must use the solution (44) in terms of the Mittag-Leffler functions, first by introducing at $\gamma_\varepsilon \Gamma(1+\varepsilon)t^{1+\varepsilon}/M \gg 1$ the asymptotic behavior of these functions. This allows expressing the MSD as

$$X(t) \approx \frac{2k_B T}{M\omega^2}\left[1 - E_{1-\varepsilon}\left(-\frac{\omega^2 M t^{1-\varepsilon}}{\gamma_\varepsilon \Gamma(1+\varepsilon)}\right)\right], \tag{56}$$

where $E_\alpha(z) = E_{\alpha,1}(z)$ is a one-parameter Mittag-Leffler function. At $\varepsilon = 0$, by using $E_1(z) = \exp(z)$, it follows from this equation the MSD within the standard Langevin model, $X(t) \approx 2k_B T[1 - \exp(-\omega^2 Mt/\gamma)]/M\omega^2$, $\gamma t/M \gg 1$ [95]. At $\omega \to 0$ this gives the correct behavior at long times, $X(t) \approx 2k_B Tt/\gamma$. For arbitrary $\varepsilon$ and $M\omega^2 t^{1-\varepsilon} \ll \gamma_\varepsilon \Gamma(1+\varepsilon)$, which is possible for long times but small $\omega$, the expansion $E_\alpha(z) = \sum_{j=0}^\infty z^j \Gamma^{-1}(\alpha j+1)$ must be used in (56), which gives

$$X(t) = \frac{2k_B T}{\gamma_\varepsilon \Gamma(1+\varepsilon)\Gamma(2-\varepsilon)}\left(t^{1-\varepsilon} - \frac{M\omega^2 \Gamma(2-\varepsilon)}{\gamma_\varepsilon \Gamma(1+\varepsilon)\Gamma(3-2\varepsilon)}t^{2-2\varepsilon} + ...\right). \tag{57}$$

In what follows we will refer to this case as to the weak trap approximation, while Eq. (55) describes the MSD for a particle in a "strong trap".

By using Eq. (55), we obtain the strong-trap $\langle\phi^2(t)\rangle$ for (2) and (8) that describes the attenuation in the steady-gradient experiment,

$$\langle\phi^2(t)\rangle \approx \frac{2k_B T\gamma_n^2 g^2}{M\omega^2}\left(\frac{t^2}{2} - \frac{\varepsilon}{\varepsilon+1}\frac{\gamma_\varepsilon}{M\omega^2}t^{\varepsilon+1}\right). \tag{58}$$

The weak trap approximation (57) gives

$$\left\langle \phi^2(t) \right\rangle = \frac{2k_B T \gamma_n^2 g^2}{\gamma_\varepsilon \Gamma(1+\varepsilon)\Gamma(2-\varepsilon)} \left( \frac{t^{3-\varepsilon}}{3-\varepsilon} - \frac{M\omega^2 \Gamma(2-\varepsilon)}{\gamma_\varepsilon \Gamma(1+\varepsilon)\Gamma(3-2\varepsilon)} \frac{t^{4-2\varepsilon}}{4-2\varepsilon} + ... \right). \tag{59}$$

The echo long-time approximation for the strong trap is obtained from Eq. (22),

$$\left\langle \phi^2(\delta, \Delta) \right\rangle \approx -2k_B T \left( \frac{\gamma_n g}{M\omega^2} \right)^2 \varepsilon \gamma_\varepsilon \psi_{\varepsilon-1}, \tag{60}$$

where $\psi_\alpha = [(\Delta+\delta)^{\alpha+2} + (\Delta-\delta)^{\alpha+2} - 2\Delta^{\alpha+2} - 2\delta^{\alpha+2}][(\alpha+1)(\alpha+2)]^{-1}$. At $\varepsilon = 0$ (the standard Langevin case)

$$\left\langle \phi^2(\delta) \right\rangle \approx \frac{4k_B T \gamma \gamma_n^2 g^2}{(M\omega^2)^2} \delta \tag{61}$$

and does not depend on $\Delta$. For the steady-gradient echo at $t = 2\tau$ ($\delta = \Delta = \tau$)

$$\left\langle \phi^2(2\tau) \right\rangle \approx \frac{8(1-2^{\varepsilon-1})\gamma_\varepsilon}{1+\varepsilon} \cdot \frac{k_B T \gamma_n^2 g^2}{(M\omega^2)^2} \tau^{1+\varepsilon} \tag{62}$$

and at $\delta \ll \Delta$

$$\left\langle \phi^2(\delta) \right\rangle \approx \frac{4k_B T \gamma_\varepsilon \gamma_n^2 g^2}{(M\omega^2)^2 (1+\varepsilon)} \delta^{1+\varepsilon}. \tag{63}$$

The influence of the weak trap is given by the second term in $X(t)$, Eq. (57). The full result is

$$\left\langle \phi^2(\delta, \Delta) \right\rangle = \frac{2k_B T \gamma_n^2 g^2}{\gamma_\varepsilon \Gamma(1+\varepsilon)\Gamma(2-\varepsilon)} \left[ \psi_{1-\varepsilon} - \frac{M\omega^2 \Gamma(2-\varepsilon)}{\gamma_\varepsilon \Gamma(1+\varepsilon)\Gamma(3-2\varepsilon)} \psi_{2-2\varepsilon} + ... \right], \tag{64}$$

When $\varepsilon = 0$, $\left\langle \phi^2(\delta, \Delta) \right\rangle = 2k_B T \gamma_n^2 g^2 \gamma^{-1} [\psi_1 - M\omega^2 \psi_2 / 2\gamma + ...]$ with $\psi_1 = \delta^2(\Delta - \delta/3)$ and $\psi_2 = (\delta\Delta)^2$. The first term, as Eq. (52) at this limit, corresponds to the well-known result [3].

The formulas for the NMR signal attenuation based on the solutions of the GLE describing the fractional BM generalize the results from the standard Langevin theory for both free particles and particles trapped in a harmonic potential. At long times, which are of main interest from the point of view of experiments, the free particles within this model undergo subdiffusion, and the MSD of the trapped particles is, see Eq. (56), $X(t) \approx 2k_B T / M\omega^2$. In the latter case, as it is seen from Eq. (58), the NMR signal in the steady gradient experiment can be significantly influenced by the particle motion, while such $X(t)$ (time-independent) does not affect the signal of the pulsed-gradient echo at all.

In what follows we describe the experiments in which subdiffusion of water in human brain tissues [42] has been clearly demonstrated. The fractional BM thus seems to be one of the possible models to be used in interpretation of these experiments. The choice of this system is important also because of the fact that the diffusion in neuronal tissues has been associated with alterations in physiological and pathological states and its understanding can have important implications.

By using Eq. (43) [42] (see above the text after Eq. (52)), in [42] Eq. (44) was at $\Delta = \delta = t$ used in the form

$$\ln[S(t)/S(0)] = 2[(3-\varepsilon)\Gamma(2-\varepsilon)]^{-1}\gamma_n^2 g^2 D t^{3-\varepsilon}. \tag{65}$$

The correct expression should however read, from Eq. (52),

$$-\ln[S(t)/S(0)] \approx 4(2^{1-\varepsilon}-1)[(2-\varepsilon)(3-\varepsilon)\Gamma(2-\varepsilon)\gamma_\varepsilon]^{-1}\gamma_n^2 g^2 k_B T t^{3-\varepsilon}. \tag{66}$$

Along with a different non-dimensional coefficient, (65) contains the parameter $D$, which is not the diffusion coefficient of the dimension $L^2 T^{-1}$ as in [42], but should be $D_\varepsilon = k_B T / \gamma_\varepsilon$, which dimension is $L^2 T^{-1+\varepsilon}$. The determination of $D$ and $\alpha = 1 - \varepsilon$ in [42] from experiments is thus flawed. So, even if $\alpha$ was correctly extracted from the experiments, the numerical value of $D$ found from the measured $\ln[S(t)/S(0)]$ would have to be divided by $2(2^\alpha - 1)/(\alpha + 1)$, which for $\alpha \in [0, 1]$ changes from 0 to 1 (e.g., if $\alpha = 0.5$, see Table II [42], $D_\varepsilon$ should be about 1.8 times larger than $D$). The parameters $\alpha$ and $D_\varepsilon$ of the fractional BM model are easily obtainable from the comparison of the presented theory with experimental data. Such a comparison is illustrated by Fig. 4 that shows the time dependence of $S(t)$ calculated from Eqs. (66) and (65) [75] for the same $\varepsilon$ as in [42].

The use of both (65) and (66) lead to a good agreement with the experiment [42], but, for a given $\varepsilon$, at very different values of $D_\varepsilon$. If we assume that the numerical value of $D_\varepsilon$ was correctly determined in [42], an agreement with the experiment is reached for a very distinct $\varepsilon$. For example, if $(\gamma_n g)^2 D_\varepsilon = 3.48 \times 10^4$ s$^{-3+\varepsilon}$, $\varepsilon$ would be approximately 0.42 instead of 0.31 found in [42]. Note that the fit to the experimental data performed in [42] was very accurate (the chi-squared values for goodness of the fit were $< 10^{-5}$). On the other hand, our calculations represented by the dashed line in Fig. 4 very well correspond to this fit: when the difference between the fit (and thus also the data [42]) and our calculation is maximal, the relation of $S(t)$ to that from [42] is 1.01. For all the experimental times our calculations thus show a very good agreement with the data.

We are not aware about a suitable experiment on trapped particles in a fractal environment. For such particles we thus give only an illustration how the particle motion at long times (when $\gamma_\varepsilon \Gamma(\varepsilon+1) t^{\varepsilon+1} / M \gg 1$) influences the steady-gradient echo signal in the cases of strong and weak traps, which requires the conditions $M\omega^2 t^{1-\varepsilon} \gg \gamma_\varepsilon \Gamma(1+\varepsilon)$ and $M\omega^2 t^{1-\varepsilon} \ll \gamma_\varepsilon \Gamma(1+\varepsilon)$ to be fulfilled, respectively. A problem arises that for such illustration we need to know the generalized friction coefficient $\gamma_\varepsilon$ with $\varepsilon \neq 0$. This can be solved by introducing the following relation between $\gamma_\varepsilon = k_B T / D_\varepsilon$ and the usual friction coefficient $\gamma = k_B T / D$ when $\varepsilon = 0$:

$$\gamma_\varepsilon = (k_B T)^\varepsilon R^{-2\varepsilon} \gamma^{1-\varepsilon}. \tag{67}$$

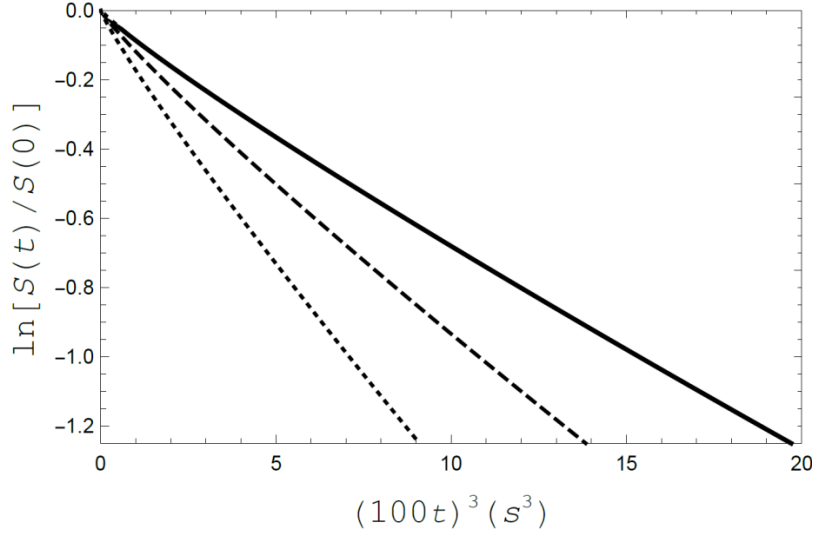

**Figure 4.** Attenuation function for the steady-gradient spin-echo signal from water molecules anomalously diffusing in a human brain tissue. The parameters correspond to the experiment ([42], Table I, Fig. 1) and $\varepsilon = 0.31$. The dashed line is calculated from Eq. (65) with $(\gamma_n g)^2 D_\varepsilon = 3.48 \times 10^4$ s$^{-3+\varepsilon}$ and very well corresponds to the experiment [42] and to Eq. (66), but if $D_\varepsilon$ is $(2-\varepsilon)/[2(2^{1-\varepsilon}-1)] \doteq 1.38$ times larger. Other lines illustrate the time dependence of the attenuation from Eq. (66) for $(\gamma_n g)^2 D_\varepsilon = 3.48 \times 10^4$ s$^{-3+\varepsilon}$ (full line) and $(\gamma_n g)^2 D_\varepsilon = 2 \times 3.48 \times 10^4$ s$^{-3+\varepsilon}$ (dotted line).

This relation or the equivalent one, $D_\varepsilon = D^{1-\varepsilon} R^{2\varepsilon}$, can be obtained by dimensional consideration. Of course, it is not the most general one. One could assume that $D_\varepsilon = a_\varepsilon + b_\varepsilon D^{1-\varepsilon} R^{2\varepsilon}$, where the term $a_\varepsilon$ has the dimension L$^2$ T$^{-1+\varepsilon}$ and $b_\varepsilon$ is dimensionless. Both these quantities characterize the fractal environment for the Brownian particle and do not depend on its radius $R$. Let us assume that $R$ decreases. The second term in $D_\varepsilon$, which is proportional to $R^{3\varepsilon-1}$ is expected to grow. This is possible if $\varepsilon < 1/3$, i.e., not $\varepsilon < 1$, as initially assumed in the fractional BM model. Moreover, with the increase of $R$ the fractal property of the environment fades away (the fractality is manifested weaker) and $D_\varepsilon$ gets closer to $D$. Thus, the term $a_\varepsilon$ can be assumed negligible and $b_\varepsilon$ equal to 1 or close to it, so that the use of (67) seems to be reasonable. Figures 5 and 6 illustrate the dependence of the attenuation function on $\varepsilon$ for the steady-gradient spin-echo signals for strong and weak traps, calculated from Eq. (60) and (64), respectively.

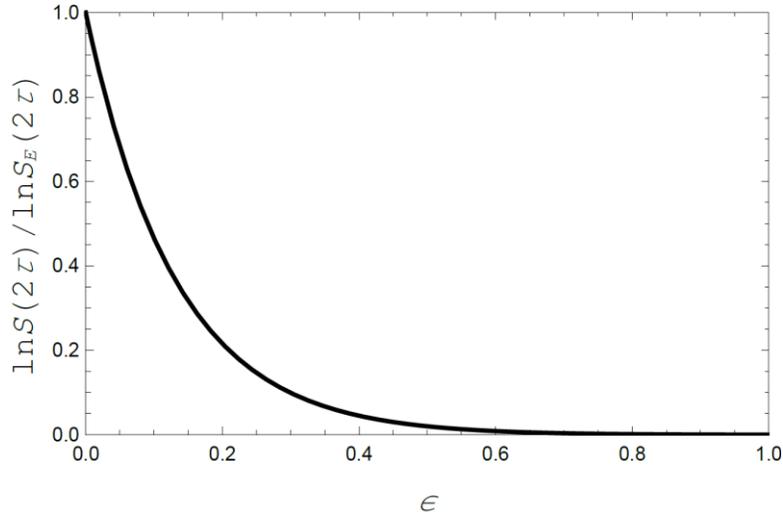

**Figure 5.** Normalized attenuation function of the steady-gradient echo signal from particles in a strong trap at long times, calculated from Eq. (60) for the model (67), as described in the text. The used parameters are $k = 10^2$ µN/m [43, 69], $\tau = 10$ ms, the viscosity is as for water at $T = 300$ K, when for particles with $R = 1$ µm the friction and diffusion coefficients are $\gamma \approx 16 \cdot 10^{-9}$ kg/s and $D \approx 0.26 \cdot 10^{-12}$ m²/s, and the mass $M \approx 4.2 \cdot 10^{-15}$ kg, assuming the particles' density close to that of water.

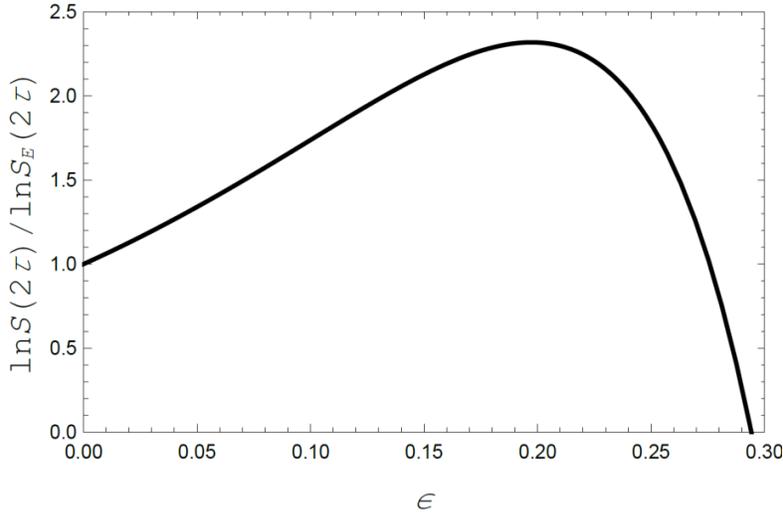

**Figure 6.** Normalized attenuation function of the steady-gradient echo signal from particles in a weak trap at long times, calculated from Eq. (64) for the model (67), as described in the text. The trap stiffness is small, $k = 1$ µN/m [96], the viscosity is as for blood at $T = 300$ K, $R = 1$ µm, $\gamma = 64 \cdot 10^{-9}$ kg/s, $D = 6.47 \cdot 10^{-14}$ m²/s, and $\tau = 10^{-2}$ s.

The results for $\ln S(2\tau)$ are normalized to $\ln S_E(2\tau)$, where $S_E(2\tau) = \exp(-2k_B T \gamma_n^2 g^2 k^{-2} \gamma \tau)$ follows coming from the Einstein theory of diffusion (see Eqs. (61) and (62) at $\varepsilon = 0$). The increase of the fractal parameter $\varepsilon$ leads to a significant difference from the attenuation in the case of normal diffusion. For the strong trap the echo signal disappears when $\varepsilon$ becomes close to 1. In Fig. 6 only small values of $\varepsilon$ are relevant, since with the growth of $\varepsilon$ the weak-trap approximation in (64) becomes inappropriate.

Finally, we note that in some works the fractional BM model is thought somewhat differently. So, in Ref. [50] the memory effects in the dynamics of protein folding was described by the model which at long times (but short enough that the influence of the potential $V(x)$ of mean force on the motion along a reaction coordinate ($x$) is negligible) shows subdiffusive behavior, $X(t) = 2Kt^\alpha$, where $0 < \alpha < 1$ and $K = C/2$ is a generalized diffusion coefficient (see Section 6 devoted to anomalous diffusion). The time evolution of the probability density of $x$, $p(x, t)$, is assumed to obey the generalized Smoluchowski equation [91, 97–99] with the time-dependent diffusion coefficient $D(t) = \alpha K t^{\alpha-1}$. This coefficient really equals to $dX(t)/2dt$, as it should be [54], but only in the long time limit. If $V(x) = kx^2/2$, the autocorrelation function (ACF) in the model [50] at long times is $\langle x(t)x(0)\rangle = (k_B T/k)\exp(-kKt^\alpha/k_B T)$. However, since the theory [50] comes from the GLE (41), the correct result for $X(t)$ is given by Eq. (56) so that the ACF should be $\langle x(t)x(0)\rangle \approx (k_B T/k)E_\alpha[-kK\Gamma(1+\alpha)t^\alpha/k_B T]$. The difference from the result [50] can be significant, as seen in Fig. 7.

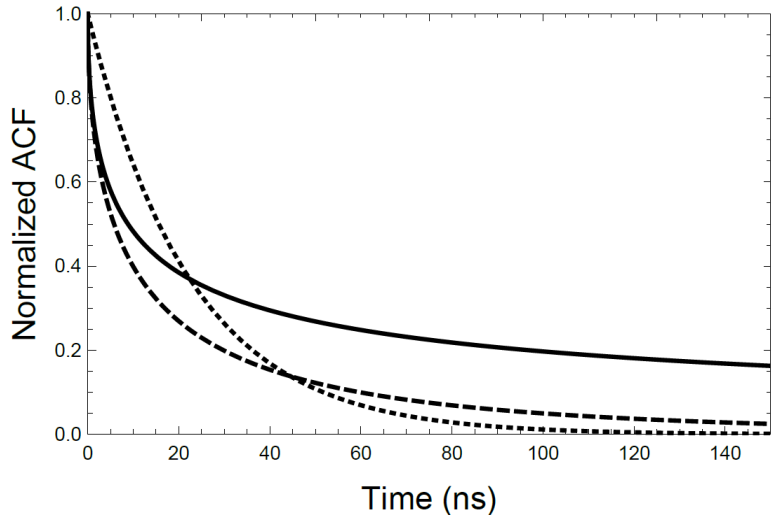

**Figure 7.** ACF for HP-35 protein in unfolded state normalized to $\langle x^2 \rangle = k_B T/k$. Full line – the fractional BM model discussed in this chapter, dashed line – the model [50], dotted line – normal diffusion model. The parameters are $k_B T/k = 0.6076$ Å$^{-2}$, $K = 0.0134$ Å$^2$ps$^{-\alpha}$, and $\alpha = 0.5132$ [100].

## CONCLUSION

In the last decade many experiments have proven that the standard Einstein and Langevin theories fail to describe the observations of the stochastic motion of particles in various systems, except for long times when the particles are in the diffusion regime. At the same time, theoretical descriptions of various NMR experimental methods, which are for more than half a century successfully applied to study the diffusion and self-diffusion processes, usually do not go beyond the long-time limit and the memoryless description of the dynamics of spin-bearing particles. A few attempts based on the generalized Langevin equation to take into account the memory effects that are revealed at shorter times were not successful. As we have shown in this chapter, either this equation was not solved correctly or the calculations of the attenuation function $S(t)$ of the NMR signals are mistaken. In the present work, the attenuation function was evaluated for two frequent examples of the NMR experiments: when the nuclear induction signal is measured in the presence of a field gradient, and for the steady and pulsed gradient Hahn spin echo. The observed attenuation is calculated through the accumulation of the spin phases in the frame rotating with the resonance frequency. Coming from the changes of the phases during the time of observation, this accumulation is represented through the mean square displacement of spins in stationary and Gaussian random processes. Several new formulas have been obtained that are valid for any times of measurements. At long times they give known results in the case of normal diffusion and describe the influence of the signal by anomalous diffusion, but are equally applicable to the stochastic motion described by other models, e.g., by various generalizations of the Langevin theory of the Brownian motion. We have considered in detail several models based on the generalized Langevin equation with memory. The corresponding basic equations of the model are solved by a unique method possessing in a very simple way the mean square displacement and other time correlation functions to be used in the interpretation of experiments. For the exponentially correlated (Ornstein-Uhlenbeck) noise when the memory in the system also exponentially decays in time the new attenuation functions are presented. It is important that while formally at long times these results coincide with the results based on the standard Langevin model, in fact the difference is significant, since the difference between $S(t)$ following from the generalized Langevin model and the model of normal diffusion is notable for much shorter experimental times. This can be used in interpretation of experiments on such systems as Maxwell (viscoelastic) fluids. Completely new results are presented coming from the natural model of hydrodynamic Brownian motion. It is shown that the diffusion limit of the attenuation functions within this model is reached very slowly, which could have an impact on the determination of the diffusion coefficients of particles, even if they are extracted from measurements with duration of gradient pulses much larger than the characteristic times of the loss of memory of the Brownian particles. In both these models at long times the particles approach the Einstein diffusion regime. The presented theory is however applicable also in situations when the diffusion is anomalous, as it is, e.g., in the case of the Brownian motion in fractal environment. An important example of a model describing such a situation is a popular model of the fractional Brownian motion. We have considered it in great details for both free particles and particles trapped in a harmonic well. The found attenuation function agrees very well with experiments carried out on human neuronal tissues. We believe that the presented results and the used approach itself could find application in interpretation of the NMR experiments on transport in various heterogeneous systems such as the network of pores filled by a solvent, anomalous Brownian motion in microemulsions and polymers, chemical reaction kinetics, distance fluctuations of particles in complex fluids, dynamics of random-coil macromolecules in gels, anomalous Brownian motion cells, protein conformational dynamics, and others.

## Acknowledgment

This work was supported by the Scientific Grant Agency of the Slovak Republic through grant VEGA 1/0250/18.

## REFERENCES

1. Callaghan, P. T. Principles of Nuclear Magnetic Resonance Microscopy; Oxford University Press: New York, 1991.
2. Kimmich, R. NMR - Tomography, Diffusometry, Relaxometry; Springer-Verlag, Berlin, Heidelberg, 1997.
3. Kimmich, R.; Fatkullin, N. Adv. Polymer Sci. 2004, 170, 1.
4. Özarslan, E.; Basser, P. J.; Shepherd, T. M.; Thelwall, P. E.; Vemuri, B. C.; Blackband S. J. J. Magn. Res. 2006, 183, 315.
5. Momot, K. I.; Kuchel, P. W. Concepts in Magn. Res. 2006, 28 A, 249.
6. Grebenkov, D. Rev. Mod. Phys. 2007, 79, 1077.
7. Grebenkov, D. J. Magn. Res. 2010, 205, 181.
8. Grebenkov, D. J. Magn. Res. 2011, 208, 243.
9. Mitchell, J.; Chandrasekera, T. C.; Gladden, L. F. J. Chem. Phys. 2013, 139, 074205.
10. Callaghan, P. T. Translational Dynamics and Magnetic Resonance: Principles of Pulsed Gradient Spin Echo NMR, Oxford University Press, Oxford, 2014.
11. Nguyen, D. V.; Grebenkov, D.; Bihan, D.; Li, J.-R. J. Magn. Res. 2015, 252, 103.
12. Singer, P. M.; Mitchell, J.; Fordham, E. J. J. Magn. Res. 2016, 270, 98.
13. Momot, K. I.; Holzapfel, N. P.; Loessner, D. Biomed. Spectroscopy and Imaging 2016, 5, 41.
14. Coffey, W. T.; Kalmykov, Yu. P.; Waldron, J. T. The Langevin Equation. With Applications to Stochastic Problems in Physics, Chemistry and Electrical Engineering; World Scientific: New Jersey, 2005.
15. Klages, R.; Radons, G.; Sokolov, I. M. (Eds.) Anomalous Transport: Foundations and Applications; Wiley-VCH: Berlin, 2008.

16. Mazo, R M. Brownian Motion. Fluctuations, Dynamics, and Applications; Oxford University Press: New York, 2009.
17. Tóthová, J.; Lisý, V. Unusual Brownian Motion, In Statistical Mechanics and Random Walks; A. Skogseid and V. Fasano, Eds.; Nova Science: New York, 2012; pp. 39-63.
18. Sutherland, W. Phil Mag 1905, 9, 781.
19. Einstein, A. Ann Phys 1905, 17, 549.
20. Einstein, A. Ann Phys 1906, 19, 289.
21. Langevin, P. C. R. Acad. Sci. (Paris) 1908, 146, 530.
22. Uhlenbeck, G. J.; Ornstein, L. S. Phys Rev 1930, 36, 823.
23. Chandrasekhar, S. Rev Mod Phys 1943, 15, 1.
24. Lisý, V.; Tóthová, J. Phys. Rev. Lett. 2017, 117, 249701.
25. Lisý, V.; Tóthová, J.  arXiv:1312.03334 [cond-mat.stat-mech]
26. Lisý, V.; Tóthová, J. Magn. Res. 2017, 276, 1.
27. Lisý, V.; Tóthová, J. Mol. Liq. 2017, 234, 182.
28. Tóthová, J.; Lisý, V. Acta Phys. Pol. 2017, 131, 1111.
29. Lisý, V.; Tóthová, J. Physica A 2018, 494, 200.
30. Stepišnik, J. Physica B 1994, 198, 299.
31. Callaghan, P. T.; Stepišnik, J. Adv. Magn. Opt. Res. 1996, 19, 325.
32. Torrey, H. C. Phys. Rev. 1956, 104 563.
33. Douglas, D. C.; McCall, D. W. J. Phys. Chem. 1958, 62, 1102.
34. Kärger, J.; Vojta, G. Chem. Phys. Lett. 1987, 141, 411.
35. Kärger, J.; Pfeifer, H.; Vojta, G. Phys. Rev. A 1988, 37, 4514.
36. Widom, A.; Chen, H. J. J. Phys. A: Math. Gen. 1995, 28, 1243.
37. Fan, Y.; Gao, J.-H. Phys. Rev. E 2015, 92, 012707.
38. Jarenwattananon, N. N.; Bouchard, L.-S. Phys. Rev. Lett. 2015, 114, 197601. Erratum: Phys. Rev. Lett. 2016, 116, 219903.
39. Laun, F. B.; Müller, L.; Kuder, T. A. Phys. Rev. E 2016, 93, 032401.
40. Zänker, P. P.; Schmidt, J.; Schmiedeskamp, J.; Acosta, R. H.; Spiess, H. W. Phys. Rev. Lett. 2007, 99, 263001.
41. Zänker, P. P.; Schmidt, J.; Schmiedeskamp, J.; Acosta, R. H.; Spiess, H. W. Chem. Phys. Lett. 2009, 481, 137.
42. Cooke, J. M.; Kalmykov, Yu. P.; Coffey, W. T.; Kerskens, Ch. M. Phys. Rev. E 2009, 80, 061102.
43. Franosch, Th.; Grimm, M.; Belushkin, M.; Mor, F. M.; Foffi, G.; Forró, L.; Jeney, S. Nature 2011, 478, 85.
44. Li, T.; Raizen, M. Ann. Phys. (Berlin) 2013, 525, 281.
45. Kheifets, S.; Simha, A.; Melin, K.; Li, T.; Raizen, M. G. Science 2014, 343, 1493.
46. Mori, H. Prog. Theor. Phys. 1965, 33, 423.
47. Kubo, R. Rep. Prog. Phys. 1966, 29, 255.
48. Grimm, M.; Jeney, S.; Franosch, Th. Soft Matter 2011, 7, 2076.
49. Lisý, V.; Tóthová, J. Transport Theory and Stat. Phys. 2013, 42, 365.

50. Satija, R.; Das, A.; Makarov, D. E. J. Chem. Phys. 2017, 147, 152707.
51. Kubo, R.; Tomita, K. J. Phys. Soc. Jpn. 1954, 9, 888.
52. Nørrelykke, S. F.; Flyvberg, H. Phys. Rev. E 2011, 83, 041103.
53. Ferreira, R. M. S.; Santos, M. V. S.; Donato, C. C.; Andrade, J. S., Jr.; Oliveira, F. A. Phys. Rev. E 2012, 86, 021121.
54. Tóthová, J.; Vasziová, G.; Glod, L.; Lisý, V. Eur. J. Phys. 2011, 32, 645.
55. Tóthová, J.; Vasziová, G.; Glod, L.; Lisý, V. Eur. J. Phys. 2011, 32, L47.
56. Vladimirsky, V. Zhur. Eksp. Teor. Fiz. 1942, 12, 199.
57. Vladimirsky, V.; Terletzky, Ya. Zhur. Eksp. Teor. Fiz. 1945, 15, 259.
58. Lisý, V.; Tóthová, J. arXiv:cond-mat/0410222.
59. Lisý, V.; Tóthová, J.; Glod, L. Int. J. Thermophys. 2013, 34, 629.
60. Tóthová, J.; Lisý, V. Acta Phys. Slov. 2015, 65, 1.
61. Tóthová, J.; Lisý, V. Phys. Lett. A 2016, 380, 2561.
62. Hirschfelder, J. O.; Curtis, C. F.; Bird, R. B. The Molecular Theory of Gases and Liquids, Wiley, New York, 1964.
63. Hahn, E. L. Phys. Rev. 1950, 80, 580.
64. Stejskal, E. O.; Tanner, J. E. J. Chem. Phys. 1965, 42, 288.
65. Jug, G. Chem. Phys. Lett. 1986, 131, 94.
66. Lukić, B.; Jeney, S.; Tischer, C.; Kulik, A.J.; Forró, L.; Florin, E.-L. Phys. Rev. Lett. 2005, 95, 160601.
67. Lukić, B.; Jeney, S.; Sviben, Z.; Kulik, A. J.; Florin, E.-L.; Forró, L. Phys. Rev. E 2007, 76, 011112.
68. Li, T.; Kheifets, S.; Medellin, D.; Raizen, M. G. Science 2010, 328, 1673.
69. Huang, R.; Chavez, I.; Taute, K. M.; Lukić, B.; Jeney, S.; Raizen, M.; Florin, E.-L. Nature Phys. 2011, 7, 576.
70. Pusey, P. N. Science 2011, 332, 802.
71. Boussinesq, J. C. R. Acad. Sci. (Paris) 1885, 100, 935.
72. Basset, A. B. Phil. Trans. R. Soc. A 1888, 179, 43.
73. Landau, L. D.; Lifshitz, E. M. Hydrodynamics; Nauka: Moscow, 1986.
74. Abramowitz, A; Stegun, I. A. Handbook of Mathematical Functions; National Bureau of Standards: Washington, DC, 1964.
75. Wolfram Research, Inc., Mathematica, Version 11, Champaign, IL, 2017.
76. Mandelbrot B. B.; Van Ness, J. W. SIAM Review 1968, 10, 422.
77. Metzler, R.; Klafter, J. Phys. Rep. 2000, 339, 1.
78. Lutz, E. Phys. Rev. E 2001, 64, 051106.
79. Viñales, A. D.; Despósito, M. A. Phys. Rev. E 2006, 73, 016111.
80. Despósito, M. A.; Viñales, A. D. Phys. Rev. E 2009, 80, 021111.
81. Goychuk, I. Phys Rev E 2009, 80, 046125.
82. Tarasov, V. E.; Tarasova, V. V. Ann. Phys. 2017, 383, 579.
83. Karaman, M. M.; Wang, H.; Sui, Y.; Engelhard, H. H.; Li, Y.; Zhou, X. J. NeuroImage:

Clinical 2016, 12, 707.
84. Panja, D. J. Stat. Mech.: Theory and Experiment (JSTAT) 2010, L02001.
85. Panja, D. J. Stat. Mech.: Theory and Experiment (JSTAT) 2010, P06011.
86. Lisý, V.; Tóthová, J.; Brutovsky, B.; Zatovsky, A.V. The dynamics of polymers in solution with hydrodynamic memory. In Soft Condensed Matter: New Research; K.I. Dillon; Ed.; Nova Science: New York, 2007; pp. 35-77.
87. Tóthová, J.; Lisý, V.; Zatovsky, A. V. J. Chem. Phys. 2003, 119, 13135.
88. Lisý, V.; Tóthová, J.; Zatovsky, A. V. J. Chem. Phys. 2004, 121, 10699.
89. Lisý, V.; Tóthová, J.; Zatovsky, A. V. J. Stat. Mech.: Theory and Experiment (JSTAT) 2008, P01024.
90. Talkner, P.; Braun, H.-B. J. Chem. Phys. 1988, 88, 7537.
91. Chaudhury, S.; Cherayil, B. J. J. Chem. Phys. 2006, 125, 184505.
92. Sevilla, P. J.; Kenkre, V. M.; J. Phys.: Condens. Matter 2007, 19, 065113.
93. Callaghan, T.; Pinder, D. N. Macromolecules 1980, 13, 1085.
94. Piana, S.; Lindorff-Larsen, K.; Shaw, D. E. PNAS 2012, 109, 17845.
95. Glod, L.; Vasziová, G.; Tóthová, J.; Lisý, V. J. Electr. Eng. 2012, 63, 53.
96. Huang, R. Brownian Motion at Fast Time Scales and Thermal Noise Imaging; Ph.D. Dissertation, The University of Texas at Austin, 2008.
97. Okuyama, S.; Oxtoby, D. W. J. Chem. Phys. 1986, 84, 5824.
98. Okuyama, S.; Oxtoby, D. W. J. Chem. Phys. 1986, 84, 5830.
99. Gudowska-Nowak, E. Acta Phys. Pol. B 1994, 25, 1161.
100. Makarov, D. E. personal communication.